\documentclass[conference]{IEEEtran}
\IEEEoverridecommandlockouts
\usepackage{cite}
\usepackage{amsmath,amssymb,amsfonts}
\usepackage{algorithmic}
\usepackage{graphicx}
\usepackage{textcomp}
\usepackage{xcolor}
\usepackage{braket}
\usepackage{listings}
\usepackage{subcaption}
\usepackage{float}
\usepackage{stfloats} 
\usepackage{booktabs}
\usepackage{endnotes}

\def\BibTeX{{\rm B\kern-.05em{\sc i\kern-.025em b}\kern-.08em
    T\kern-.1667em\lower.7ex\hbox{E}\kern-.125emX}}

\begin{document}
\lstset{language=Python} 

\title{Tailored Quantum Device Calibration with Statistical Model Checking}

\author{
\IEEEauthorblockN{Filip Mazurek}
\IEEEauthorblockA{\textit{Department of Electrical and}\\
\textit{Computer Engineering} \\
\textit{Duke University}\\
Durham, NC, USA \\
filip.mazurek@duke.edu}
\and
\IEEEauthorblockN{Marissa D'Onofrio}
\IEEEauthorblockA{\textit{Department of Electrical and}\\
\textit{Computer Engineering} \\
\textit{Duke University}\\
Durham, NC, USA \\
marissa.donofrio@duke.edu}
\and
\IEEEauthorblockN{Andrew Van Horn}
\IEEEauthorblockA{\textit{Department of Electrical and}\\
\textit{Computer Engineering} \\
\textit{Duke University}\\
Durham, NC, USA \\
andrew.van.horn@duke.edu}
\and
\IEEEauthorblockN{Jiyong Yu}
\IEEEauthorblockA{\textit{Department of Electrical and}\\
\textit{Computer Engineering} \\
\textit{Duke University}\\
Durham, NC, USA \\
jiyong.yu@duke.edu}
\and
\IEEEauthorblockN{Kavyashree Ranawat}
\IEEEauthorblockA{\textit{Department of Electrical and}\\
\textit{Computer Engineering} \\
\textit{Duke University}\\
Durham, NC, USA \\
kavya.ranawat@duke.edu}
\and
\IEEEauthorblockN{Jungsang Kim}
\IEEEauthorblockA{\textit{Department of Electrical and}\\ \textit{Computer Engineering} \\
\textit{Department of Physics} \\
\textit{Duke University}\\
Durham, NC, USA \\
jungsang@duke.edu}
\and
\IEEEauthorblockN{Kenneth R. Brown}
\IEEEauthorblockA{\textit{Department of Electrical and}\\
\textit{Computer Engineering} \\
\textit{Department of Physics} \\
\textit{Department of Chemistry} \\
\textit{Duke University }\\
Durham, NC, USA \\
kenneth.r.brown@duke.edu}
}

\maketitle

\begin{abstract}

Quantum devices require precisely calibrated analog signals, a process that is complex and time-consuming.
Many calibration strategies exist, and all require careful analysis and tuning to optimize system availability.
To enable rigorous statistical evaluation of quantum calibration procedures, we leverage statistical model checking (SMC), a technique used in fields that require statistical guarantees.
SMC allows for probabilistic evaluation of properties of interest, such as a certain parameter's time to failure.
We extend the SMC for Processor Analysis (SPA) framework, which uses SMC for evaluation of classical systems, to create SPA for Quantum calibration (SPAQ) enabling simplified tuning and analysis of quantum system calibration.
We focus on a directed acyclic graph-based calibration optimization scheme and demonstrate how to craft properties of interest for its analysis.
We show how to use SPAQ to find lower bounds of time to failure information, hidden node dependencies, and parameter threshold values and use that information to improve simulated quantum system availability through calibration scheme adjustments.

\end{abstract}


\section{Introduction}
\label{sec:intro}

Determining and optimizing control parameters in quantum computing systems is a fundamental challenge, as gate and program quality is dependent on the application of carefully tuned analog signals. These signals drift over time due to environmental conditions and imperfect hardware, leading to errors in computational outcomes and the need for frequent recalibration 
\cite{DiVincenzo_2000, gottesman2002introduction, cortez2017rapid, proctor2020detecting}.
Systems often rely on manual adjustments or long, linear calibration processes which require system overhead and during which the system continues to drift \cite{wu2021strong, wang2020high, ibm_calibration}.
This drift can be nonlinear or unpredictable, and error due to drift may have hidden layers of dependencies. 

In recent years, several approaches have been introduced to reduce resource requirements for calibration of quantum systems \cite{wittler2021integrated, egger2014adaptive, jeanette2025blind, gerster2022experimental, maksymov2021optimal, ai_calibration}.
Effective implementation of calibration protocol depends on factors such as termination criteria for parameter optimization, interval frequency for calibration job submission, or dependency definitions in a calibration graph. 
Often manual thresholds and heuristics are used, or termination criteria may be chosen without full consideration of correlation between errors. 
A robust approach to analysis of calibration data and construction of calibration protocols should prioritize statistical validity while minimizing data requirements and system runtime.

In this article, we propose statistical model checking (SMC) for quantum calibration. SMC is a technique widely used in fields that depend heavily on statistical guarantees, such as hardware verification of sensitive devices~\cite{arney2010patient, kwiatkowska2018probabilistic, roohi:hscc:2017}.
We leverage the SMC for Processor Analysis (SPA) framework~\cite{spa}, originally developed to aid in classical computer processor analysis, and extend it to assist in quantum calibration. This provides a statistically rigorous approach for determining relevant properties of quantum systems, including:

\begin{itemize}
    \item Prediction of calibration experiment time to failure using statistical bounds.
    \item Calculation of confidence bounds for key calibration parameters.
    \item Identification of co-occurring failures to reveal dependencies among calibration experiments.
    \item Determination of dependent parameter thresholds required to maintain system integrity.
\end{itemize}

To examine these properties, we introduce the SPA for Quantum calibration (SPAQ) framework and apply it to analyze the Optimus calibration algorithm introduced by Kelley \textit{et. al.} \cite{google_optimus}.
SPAQ provides insight into optimal settings of Optimus hyperparameters, such as hidden connections in the calibration graph and acceptable threshold values. 

The contributions of this work are as follows:

\begin{itemize}
    \item We extend SPA for the analysis of graph-based quantum computer calibration.
    \item We develop the SPAQ framework to enable push-button analysis of calibration results.
    \item We show how SPAQ can achieve statistically rigorous results with a desired confidence level on simulated calibration nodes.
    \item We publicly distribute SPAQ for use by academia and industry.\endnote{https://github.com/filipmazurek/spa-quantum-calibration}
    
\end{itemize}

\begin{figure*}
    \centering
    \includegraphics[width=\linewidth]{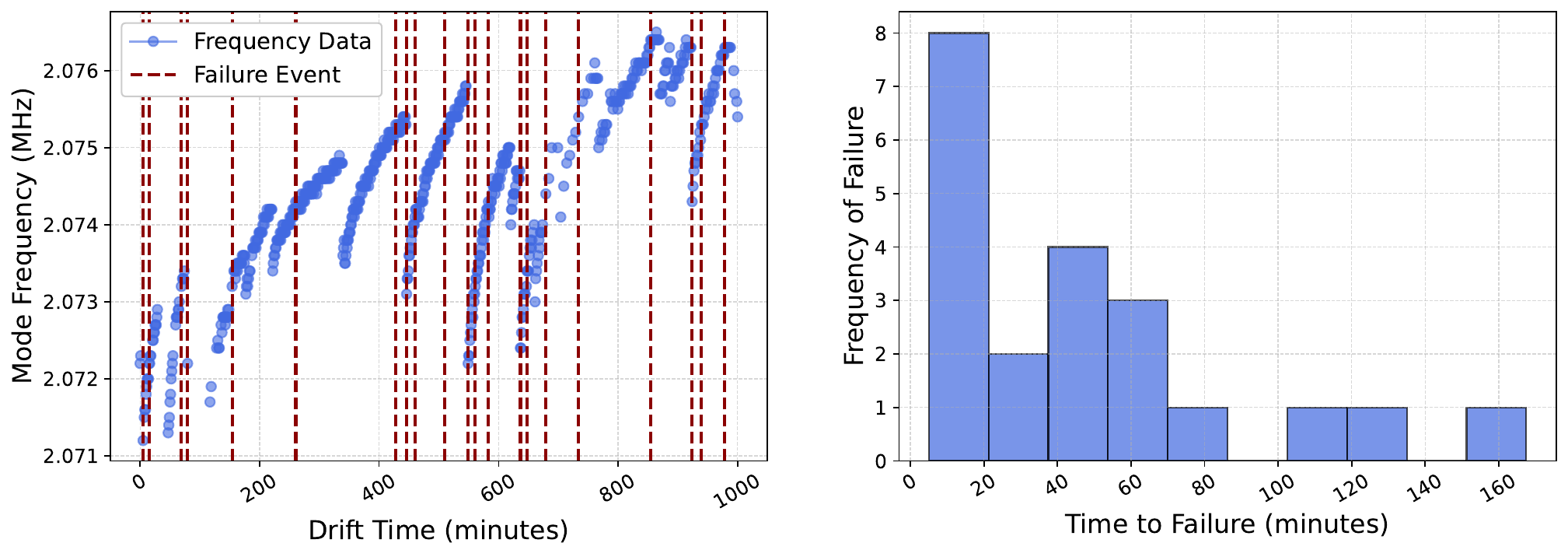}
    \caption{(Left) Motional mode frequency of a trapped $^{171}$Yb$^{+}$ ion, monitored for 18 hours. Vertical lines show simulated failure events where the mode frequency drifts more than 1 kHz. (Right) Corresponding time-to-failure histogram.}
    \label{fig:real-device-drift}
\end{figure*}

\section{Background}
\label{sec:background}

In this section, we discuss properties and effects of drift in quantum systems as well as approaches taken by state-of-the-art calibration schemes.

\subsection{Variability and Drift}
Noise and drift are fundamental challenges in quantum systems due to imperfections in control equipment as well as fluctuations in environmental factors.
Detection and suppression is inherently difficult due to sample-apparatus correlation, meaning samples may not accurately predict future observations \cite{van2013quantum}. 
This is compounded by the non-Markovian and non-Gaussian nature of many noise processes, such as dephasing noise or heating mechanisms \cite{sung2019non, fogarty2015nonexponential, brownnutt2015ion}.
Over time, parameter drift contributes to fidelity decay, compromising system performance. Consequently, characterization and calibration techniques must be designed to remain robust to these disturbances.

We illustrate parameter drift in a real system in Fig. \ref{fig:real-device-drift} (Left) by tracking the frequency of a motional mode used for gate operations in a trapped ion quantum computer \cite{spivey2021high}.
Drift in motional mode frequency may be caused by factors such as fluctuating trap voltages or charging.
If not appropriately addressed, essential gate operations will be driven at incorrect frequencies degrading system performance. 
We see that the drift rate is variable, and events such as ion reloading can create discontinuities in parameter value. 
Failure events are marked with vertical dashed lines and are defined as frequency deviations greater than 1 kHz from the most recent calibration value. 

When drift is nonlinear, system failure does not follow a predictable pattern~\cite{cortez2017rapid}. 
Linear drift would predict an approximately normal time to failure distribution, making Bayesian statistical approaches convenient to analyze and predict when next failures will occur.
We create a histogram of time to failure data for motional mode frequency drift in Fig.~\ref{fig:real-device-drift} (Right). 
Time to failure is defined as time between failure events as marked in Fig. \ref{fig:real-device-drift} (Left). 
The resulting distribution is not normal and therefore cannot be analyzed using Bayesian statistics. 
Nonlinear drift and non-Gaussian distributions, coupled with expensive and limited data collection, motivate SMC as a tool for data analysis, as detailed in  section~\ref{sec:smc}.

\subsection{Calibration Approaches}
Calibration is challenging to perform efficiently because each calibration contains many interdependent parameters and relies on the successful completion of many preliminary experiments.
Furthermore, as quantum systems scale, the calibration burden grows rapidly due to the exponential increase in number of parameters to calibrate. This includes the increasing number of qubit pairs that can be used to perform two-qubit gates, as well as an expanding array of control systems that must be precisely coordinated.
This problem is already apparent in current generation quantum systems such as IBM devices where systems larger than 7 qubits require 90 minutes of daily calibration plus 2-3 minutes of hourly calibration \cite{ibm_calibration, abughanem2025ibm}. 

Consequently, various calibration approaches have been developed to minimize calibration time and maximize system availability.
Blind calibration addresses the sample-apparatus correlation problem by determining calibration parameters without explicit knowledge of initial state preparations or detailed error models, using tomographic data for error reconstruction~\cite{jeanette2025blind}.
Bayesian calibration employs Bayesian statistical methods to iteratively refine gate parameters~\cite{gerster2022experimental}.
Other examples of calibration approaches utilize architecture-dependent properties \cite{deng2024calibration, fan2025calibrating, maksymov2021optimal}, continually calibrate during error detection~\cite{kelly2016scalable}, or adjust quantum computation through projective simulation~\cite{tiersch2015adaptive}.

We focus on the Optimus calibration framework~\cite{google_optimus}, which structures calibration procedures using a directed acyclic graph (DAG) to systematically manage dependencies and optimize parameter tuning.
Optimus is our chosen algorithm to demonstrate tuning using SMC methods due to familiarity with the calibration scheme and its implementation in our control software for real experimental systems~\cite{dax_optimus}. This ensures we can apply our SMC approach to a physical system in the future.
We describe the Optimus algorithm in Section~\ref{sec:optimus}.

{
\begin{table*}
\fontsize{8}{8}\selectfont
\resizebox{\textwidth}{!}{
\begin{tabular}{p{7cm}p{7cm}}
\toprule
\textbf{Property Template} & \textbf{Examples}  \\ 
\midrule
metric $\gtrless$ threshold  & \begin{tabular}[c]{@{}l@{}}performance $>$ A;\\ power $<$ B;\\ MTTF $>$ C;\end{tabular}     \\
\midrule
threshold1 $>$ metric $>$ threshold2    & \begin{tabular}[c]{@{}l@{}}A $>$ performance $>$ B;\\ C $>$ power $>$ D;\\ E $>$ MTTF $>$ F;\end{tabular}    \\
\midrule
\% time in state $\gtrless$ threshold   & \% time spent in check\_data $<$ A      \\
\midrule
avg \# time/event $\gtrless$ threshold        & avg time between calibration events $>$ A    \\
\midrule
metric1 $\gtrless$ threshold $\to$ metric2 $\gtrless$ threshold             & node1 failures $>$ A $\to$ node2 failures $>$ B     \\
\midrule
event1 occurs $\to$ Prob [event2 occurs within C cycles] $\gtrless$ threshold & if node1 fails check\_data, probability of node2 failing check data occurring within C cycles $<$ $\mathrm{P_B}$; immediately after calibrating node1, probability of it being checked again within C cycles $<$ $\mathrm{P_B}$ \\
\midrule
event occurs on system 1 $\to$ Prob [event occurs on system 2 within X time] $\gtrless$ threshold        & if X gate fails check\_data on system 1, probability of X gate failing check\_data on system 2 occurring within 5 minutes $<$ $\mathrm{P_B}$ \\
\bottomrule
\end{tabular}
}
\vspace{2pt}
\caption{Non-exhaustive list of properties that one can evaluate with SMC}  \label{table:properties}
\end{table*}
}

\section{Statistical Model Checking}
\label{sec:smc}
Given the nonlinear drift, non-Gaussian distributions, and high data collection overhead associated with calibration in quantum systems, we introduce SMC as a method for analysis. 
SMC is a statistical methodology that performs hypothesis testing on properties expressed in temporal logic, using defined levels of confidence and proportion.
SMC offers rigorous statistical guarantees without requiring assumptions about the underlying distribution and is particularly useful in scenarios with small samples sizes~\cite{wang:tecs:2019, zarei:hscc:2020}.
In this work we apply SMC to evaluate calibration performance using selected properties, while emphasizing extensive potential applications outside the scope of this study.

\subsection{SPA Framework for SMC}

The SPA framework was developed to simplify SMC use and enable confidence bound analysis~\cite{spa}.
SPA is tailored to provide an end-to-end solution to data analysis through hypothesis testing and calculation of upper and lower bounds for desired properties in classical computing.
A key aspect of SMC, and by extension SPA, is the use of proportion $F$, which represents the probability that a given sample satisfies the tested property.
Intuitively, it is the fraction of the population which satisfies the property.

Formally, SPA enables the evaluation of binary (true/false) properties with a specified proportion $F$ at a desired confidence level $C$.
For example, a researcher might test the hypothesis: ``The node's time-to-failure is less than 10 minutes in at least 80\% of cases, with 90\% confidence.''
A key contribution of SPA was the ability to determine confidence intervals based on properties by omitting the threshold value to test, such as: ``What is the 90\% confidence interval for the node's time-to-failure in 80\% of all cases?''~\cite{casella2021statistical}.

Properties expressed in temporal logic are not limited to metric calculations.
While SMC may be used to find a metric such as time to failure for a calibration node, its notable utility is its ability to analyze \emph{any} property expressed in temporal logic, a formal symbolic language~\cite{pnueli:focs:1977, hansson:fac:1994, maler2004monitoring}.
As the name of logic language suggests, it is useful for testing for events which occur in different points in time, such as checking if event 2 occurs within 20 cycles of event 1.
Table~\ref{table:properties} presents a number of sample property templates and how they may be used in quantum system calibration.

A thorough description of SMC, SPA, and their underpinnings may be found in~\cite{agha2018survey, legay2010statistical, spa}.

\section{The Optimus Calibration Algorithm}
\label{sec:optimus}
\begin{figure*}[!t]
    \centering
    \includegraphics[width=\linewidth]{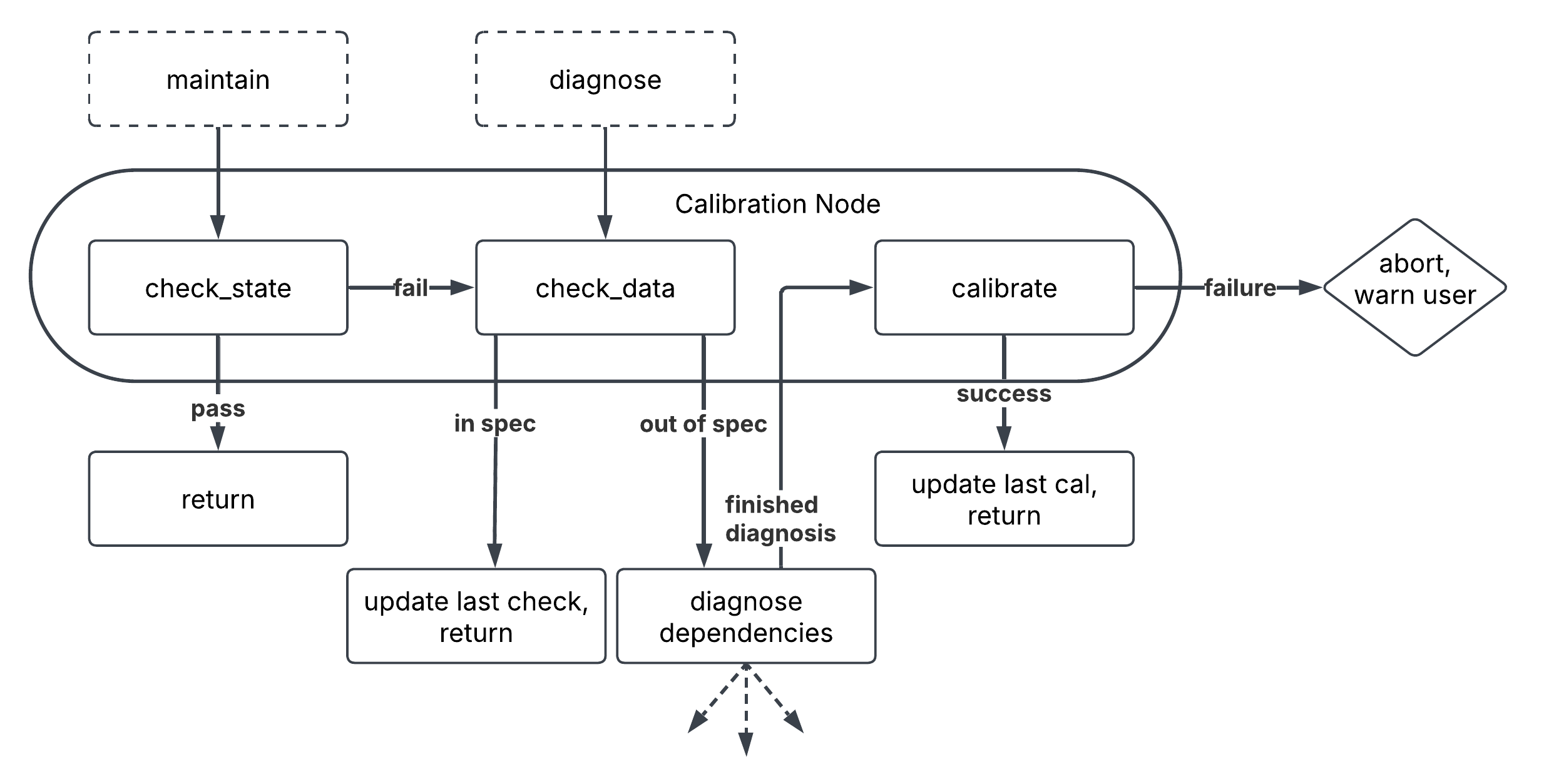}
    \caption{Internal structure of an Optimus node. Adapted from~\cite{dax_optimus}.}
    \label{fig:optimus-node}
\end{figure*}

Calibrating quantum systems involves a series of experiments to determine the parameters required for accurate operations.
As devices scale in complexity, automating the calibration process is essential to maintain quantum systems within specification.
In this work, we adopt the Optimus calibration algorithm~\cite{google_optimus} as our calibration framework.
Here we present a detailed overview of Optimus to set the stage for enhancing its performance through statistical methods.

\subsection{Algorithm Summary}
Optimus organizes calibration experiments into a DAG, where each experiment is represented as a node, and dependencies between calibrations form directed edges.
This structure allows for efficient traversal and updating of calibration states, reducing the time and complexity needed to maintain the quantum device within specification.

Fig. \ref{fig:optimus-node} depicts an Optimus node, which performs the following operations:

\begin{itemize}
    \item \texttt{check\_state}: A classical check that verifies whether the specified 
    \texttt{timeout} has elapsed (fail) or not (pass) since the last in-spec verification of the node.
    \item \texttt{check\_data}: A relatively fast experiment used to determine whether the calibration remains within a user-defined specification (in spec) or not (out of spec).
    \item \texttt{calibrate}: A comprehensive experiment which updates parameter values to optimize experimental outcome. This procedure typically requires significantly more time than \texttt{check\_data}. Failure thresholds for successful determination of calibration parameters are also user defined.
\end{itemize}

\begin{figure}
   \centering
   \includegraphics[width=\linewidth]{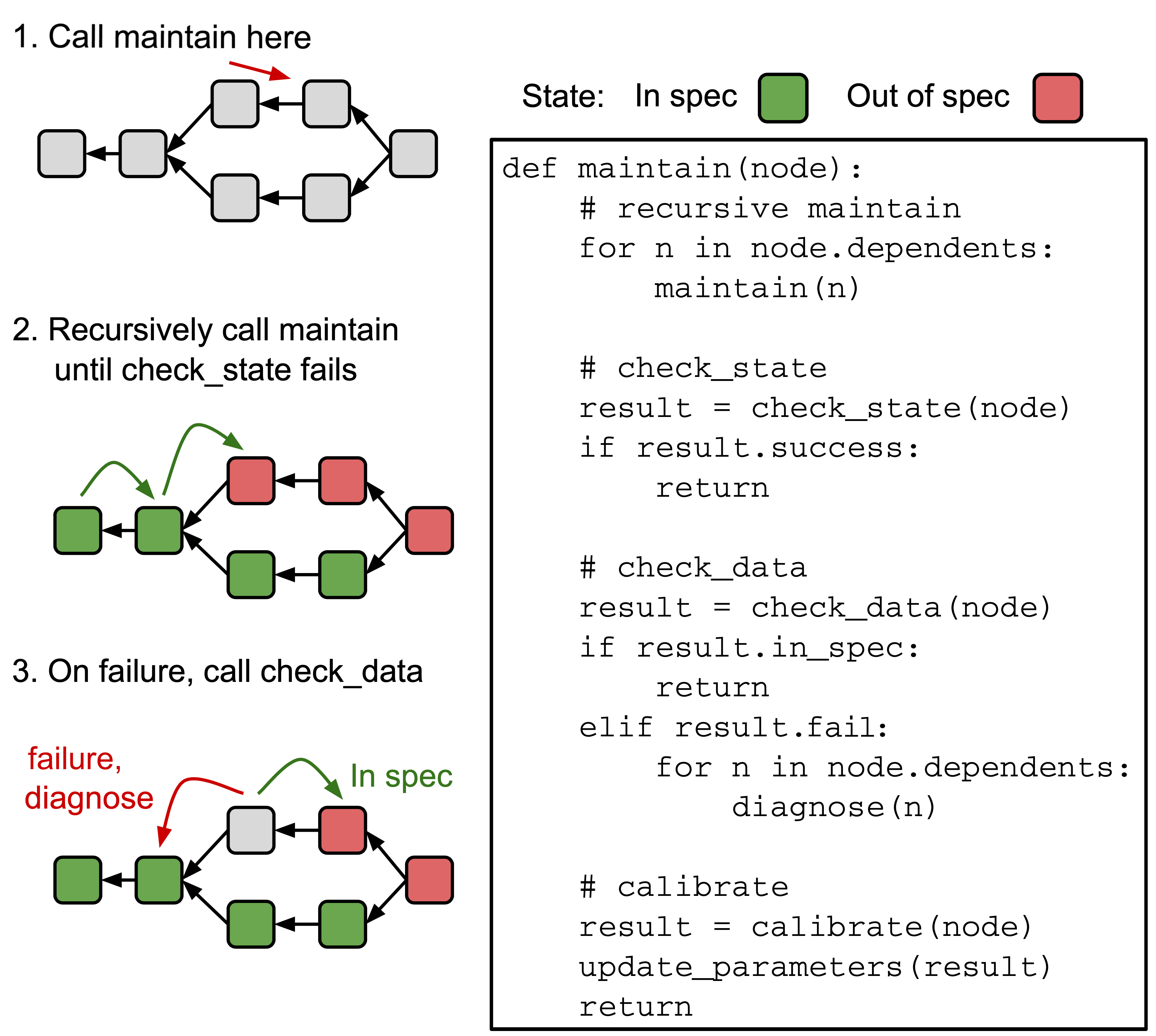}
   \caption{Optimus algorithm workflow. Adapted from~\cite{google_optimus}.} 
   \label{fig:optimus-algorithm}
\end{figure}

Each node in the Optimus DAG is connected to its direct dependencies through  edges.
When \texttt{check\_data} fails, it indicates that the experiment is out of specification.
In this case, the node's direct dependencies are checked for correctness first, as failure in the original node may stem from a miscalibrated dependency.
Optimus performs recursive checks on dependencies until the root case of the failure is identified, followed by recalibration of all failed nodes in depth-first order.
This algorithm efficiently determines the root cause of failure, and ensures that only affected nodes are recalibrated, minimizing calibration time.
The process is summarized in Fig.~\ref{fig:optimus-algorithm}.
Note that we modify the original implementation slightly:  \texttt{check\_data} can either pass or fail, and we do not differentiate between the different types of failure.

\subsection{Metaparameter Optimization}

The Optimus algorithm reduces unnecessary overhead by compartmentalizing individual calibrations.
We discuss several ways to enhance Optimus using statistical analysis, with a focus on tuning key metaparameters to ensure the highest possible quantum system availability.
The \texttt{timeout} value for each node must frequent enough to detect failures yet not so frequent as to interfere with system availability. 
Similarly, the directed edges between nodes in an Optimus graph should enable fast identification of failure root causes without unnecessary detours. 

Users may compartmentalize calibrations as a chain, define directed edges based on physical modeling of the system, or use statistical methods to determine optimal node connections.
There may be ``hidden'' factors which complicate DAG optimization, such as groups of parameters that are sensitive to environmental factors or shifts in other node parameters.
Identifying optimal node connections and faster directives for recalibration can improve quantum system availability. 
In this work, we focus on analyzing a few properties of interest, introducing statistical model checking methods to address these challenges.

\section{SPA for Quantum Systems}
\label{sec:spa-for-optimus}

As statistical model checking is not a technique familiar to most quantum computing researchers, we develop the SPA for Quantum calibration (SPAQ) framework. SPAQ is designed for statistical analysis of node-based quantum system calibration, leveraging the rigorous statistics of SMC and maintaining the ease of use of SPA.
The framework may read output data from calibration algorithms and is equipped with carefully selected SMC properties which allow insight into how the calibration procedure functions.
SPAQ is built to analyze output from Optimus using SMC.
SPAQ includes several written properties that we find useful, and is prepared for users to write custom properties to further analyze quantum system calibration procedures.
We highlight the flexibility of SMC as a core advantage: any property that can be expressed with temporal logic can be evaluated using this method.

\subsection{Properties of Interest in Quantum Calibration}

To immediately demonstrate the usefulness of SMC in quantum system calibration, we identify the following properties of interest and describe how they can be used to improve system availability:

\begin{itemize}
    \item \textbf{Node time to failure, 5th percentile}. Identifies a lower bound for node uptime following calibration. May guide adjustment of check frequency immediately following calibration.
    \item \textbf{Node time to failure, 95th percentile}. Finds the maximum expected lifetime for a node, enabling proactive checks or calibration.
    \item \textbf{Failures per time period, 95th percentile}. Detects abnormal increase in node failure, indicating potential underlying issues.
    \item \textbf{Parameter value bounds, 5th \& 95th percentile}. Defines likely thresholds for parameter values within nodes of interest. Outlier values may signal an unstable system.
    \item \textbf{Co-occurring failures}. Detects nodes which often fail together. May identify correlations between two calibration nodes assumed to be independent, suggesting hidden dependencies in the DAG.
    \item \textbf{Dependent node parameter values on failure}. Determines parameter thresholds in dependent nodes associated with downstream node failures, enabling informed parameter bound settings.
    \item \textbf{Calibration leading to failure}. Determines when a large parameter value change leads to failure of another node, inferring that the node has a large dependency on a particular parameter. May motivate a revised calibration strategy.
\end{itemize}

SPAQ comes pre-loaded with each of these parameters and enables push-button analysis of calibration results.
We interpret SPAQ results to inform modification of calibration procedures, discussed in Section~\ref{sec:evaluation}.

\subsection{SPAQ Framework}
\begin{figure}
   \centering
   \includegraphics[width=\linewidth]{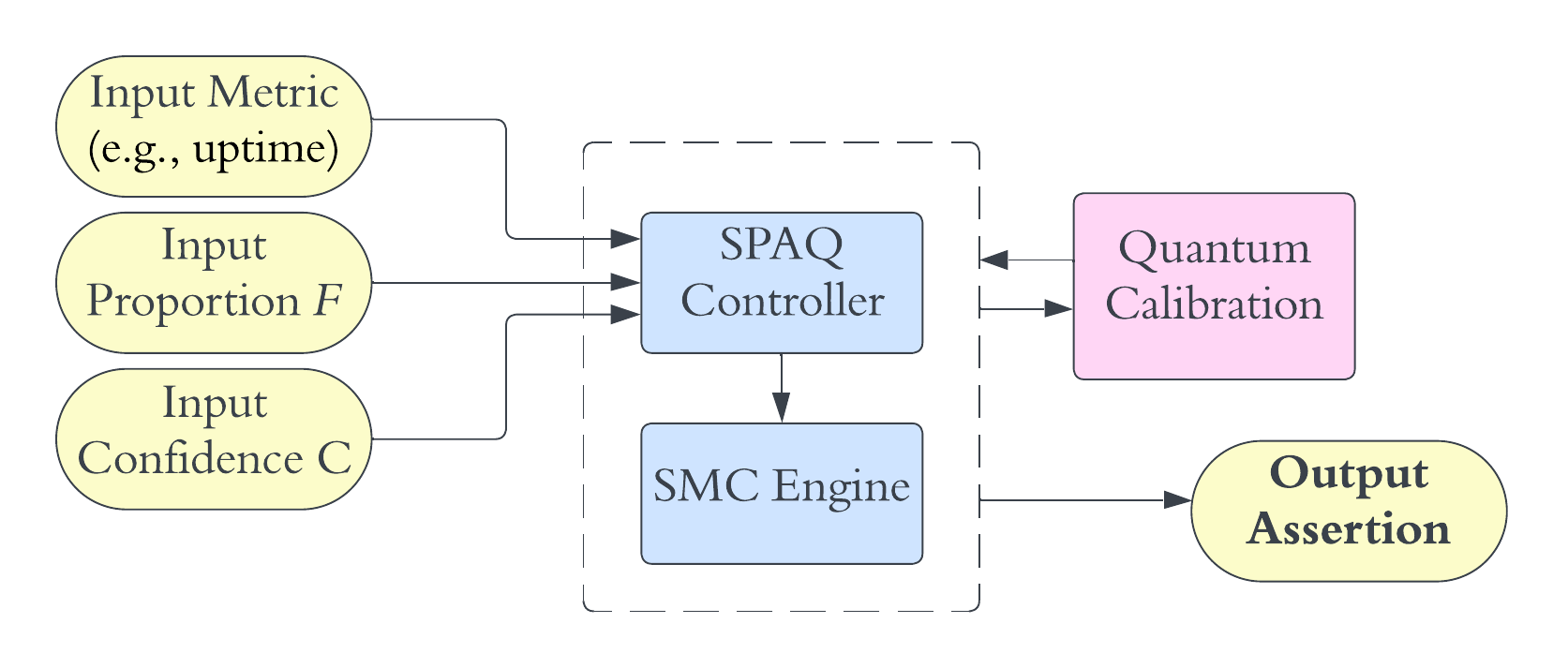}
   \caption{SPAQ structural diagram. SPAQ facilitates the coordination between quantum system data, user property specification, and the statistics calculations performed by SMC.} 
   \label{fig:spaq-structure}
\end{figure}

SPAQ coordinates calibration data from the quantum system with user-defined inputs.
The user provides the property of interest, its associated proportion and confidence level, standardized output from the calibration scheme (e.g., timestamps and the method calls), and specifies whether the output should be a binary test or confidence interval.
SPAQ automatically reads out relevant data for the property of interest and uses the SMC engine to perform all necessary calculations.
A structural diagram of the SPAQ framework is shown in Fig.~\ref{fig:spaq-structure}.
Once results are generated, it is entirely up to the user how to interpret and make use of them.
We explore possible avenues in Section~\ref{sec:evaluation}.

\section{Experimental Methodology}
\label{sec:methods}

We provide working examples of SPAQ by implementing quantum calibration experiments in simulation.
Each experiment begins by posing a concrete question about a calibration property, e.g., `Are two nodes in a calibration graph correlated?' This question is translated into a hypothesis test, such as, \emph{`Does node A fail within 15 cycles of node B with probability greater than 33\%, with at least 90\% confidence'?}
A dependency graph is then constructed with properties specifically designed to address the question. 
Most experiments are based on the dependency graph depicted in Fig. \ref{fig:optimus-dag-sim}, which is focused on calibrating a single-qubit X gate operation.

We run the Optimus graph in simulation and apply the SPAQ framework to calibration data. 
Based on the results, we modify calibration protocol and compare the updated approach against the original to evaluate improvement. 
A key metric discussed is system availability, where availability is defined as the proportion of time the system remains within specification and is not performing calibration tasks.

\subsection{Optimus Simulator}

To evaluate the effectiveness of SPAQ, we build a simulator to replicate the Optimus calibration procedure described in Section~\ref{sec:optimus}.
The simulator can be built from any number of nodes connected as an Optimus DAG.
Throughout our experiments, we use various DAG configurations, most of which are derived from the X gate calibration shown in Fig. \ref{fig:optimus-dag-sim} and described in the following section.
Nodes range in complexity from simple probabilistic failure nodes, which have a fixed chance of failure at every simulation step, to more complex nodes which utilize the ion trap gate simulator described in Kang et al.~\cite{kang2021batch} and require input from dependencies.

At every time step, parameters of each node drift according to a user-defined function, for which the SPAQ framework contains several options.
For this study, we use a logistic function: parameters remain relatively stable immediately after calibration, with drift becoming increasingly erratic over time.
Node failures are determined by their unique \texttt{check\_data} functions.
Calibration is similarly conducted: each node resets the parameter it controls to its optimal value, with some randomization.

We run 100 independent simulations of Optimus, each for 100,000 cycles, generating extensive data to evaluate SPAQ performance.
Our simulator records when the \texttt{check\_state}, \texttt{check\_data}, and \texttt{calibrate} functions are called for each node, as well as the outcome.
Furthermore, each node records its parameters both before and after calibration.
The SPAQ package includes tools for data collection during Optimus runs, while data recording within each node is written by the experimentalists creating the calibration procedure.

\subsection{Calibration Dependency Graph}

\begin{figure}
   \centering
   \includegraphics[width=0.8\linewidth]{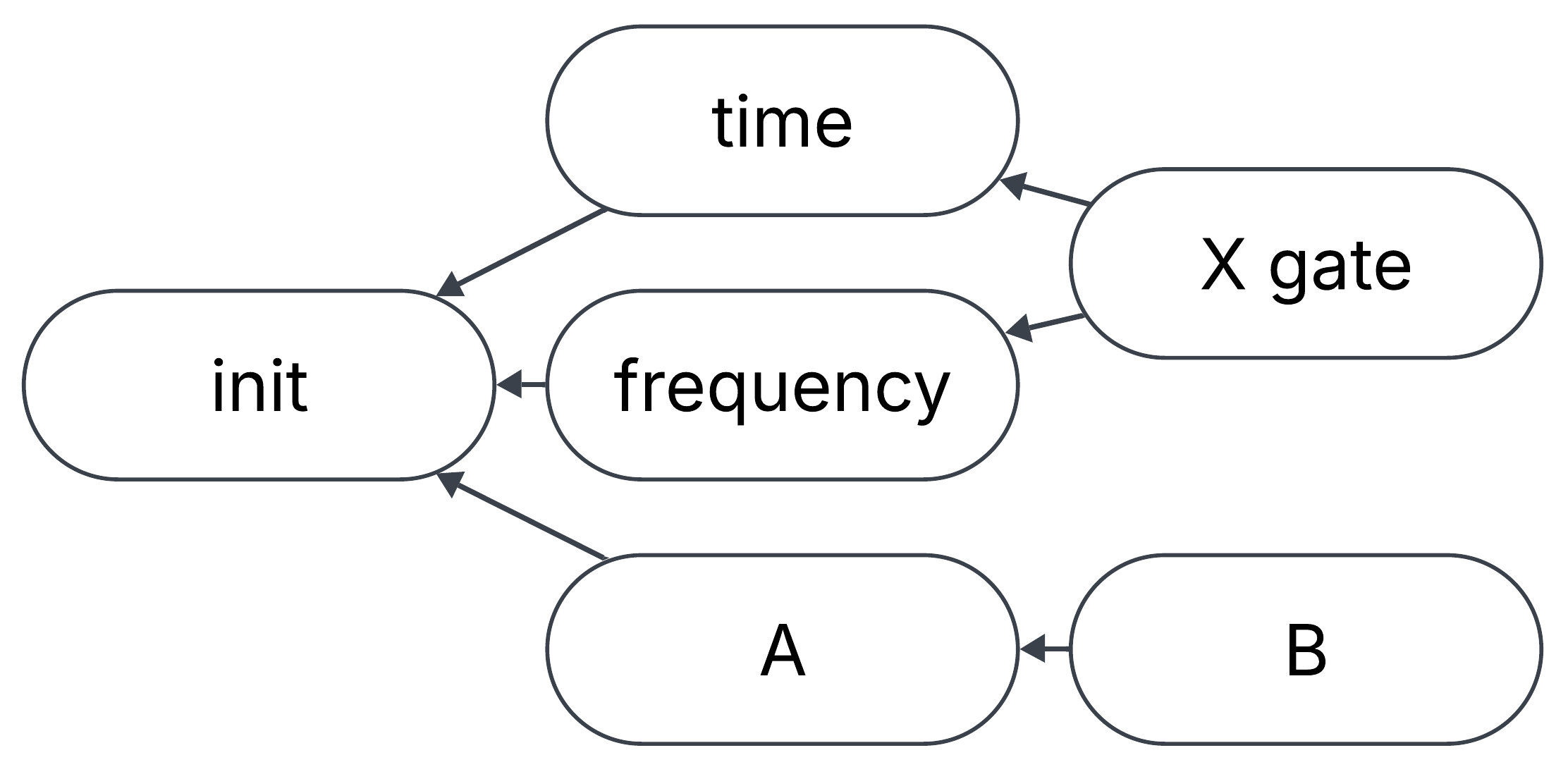}
   \caption{Directed acyclic graph representation of the Optimus graph used in simulation.} 
   \label{fig:optimus-dag-sim}
\end{figure}

We implement a reduced version of the dependency graph that would appear in a real calibration procedure for a single-qubit X gate on an ion trap quantum computer. This dependency graph, depicted in Fig. \ref{fig:optimus-dag-sim} drives the majority of our experiments.
Real calibration graphs are typically larger and more complicated; nodes not shown are assumed to behave ideally.
We focus on the main node of interest, the X gate node and its dependencies on pulse time and drive frequency, which in turn depend on state intialization.
Two generic dependency nodes A and B, are included to further demonstrate applicability of SPAQ across various internal node function.
The X gate operation is modeled using a simulator developed by Kang et al.~\cite{kang2021batch}.

The Optimus calibration graph contains the following nodes:
\begin{itemize}
    \item \textbf{X Gate Node}: Represents a single-qubit X gate, with phase as the calibration parameter. Directly depends on drive frequency and pulse time. \texttt{check\_data} evaluates X gate fidelity. 
    \item \textbf{Drive Frequency}: The tone frequency used to perform an X gate operation, found by scanning across a range of drive frequencies using a fixed pulse duration. \texttt{check\_data} verifies transition efficiency remains above a certain threshold.
    \item \textbf{Pulse Time}: Duration for which to apply the drive frequency to generate a complete transition from the $\ket{0}$ state to the $\ket{1}$ state. The time is found by scanning across a range of pulse durations for a set drive frequency. \texttt{check\_data} ensures efficient state transition.
    \item \textbf{State Initialization}: Represents qubit initialization to the $\ket{0}$state; contributes baseline noise to dependent operations. Optimal pulse time for initialization is found by scanning the pulse time to maximize population in $\ket{0}$ (i.e., dark state fidelity). \texttt{check\_data} ensures the dark state fidelity remains above a set threshold.
    \item \textbf{Nodes A and B}: Generic nodes following exponential decay functions.
    Calibration adjusts node parameters to minimize the value of interest.
    \texttt{check\_data} keeps the value of interest below a set threshold. 
    
\end{itemize}

\section{Experimental Evaluation}
\label{sec:evaluation}

We demonstrate SPAQ and its effectiveness using three experiments:

\begin{enumerate}
    \item \textbf{Delayed Checks}. Compare quantum system availability between a naive implementation of Optimus and one informed by the 5th percentile uptime data to delay initial \texttt{check\_data} calls.
    \item \textbf{Inter-Node Parameter Compensation}. Identify node failures caused by shifts in other node parameters.
    \item \textbf{Hidden Node Connections}. Find dependencies that are not explicitly defined in the Optimus graph. This includes shared effects due to environmental factors.
\end{enumerate}

\subsection{Experiment 1: Delayed Checks}
Optimus performance is expected to improve when node timeouts are carefully chosen based on relative stability of calibrations.
The most straightforward use of SPAQ is to identify metrics of interest such as as node time to failure, which can be used to set timeout intervals. 
We run Optimus on the dependency graph in Fig.~\ref{fig:optimus-dag-sim} using high-frequency checks, then use SPAQ results to appropriately delay timeout intervals after calibration.

To estimate the time to failure of the first experiment, we use SPAQ with the confidence interval calculation: \emph{What is the 95\% confidence interval for the time to failure metric at the 5th percentile ($F=0.05$)?} We apply it to each node in our calibration graph.
Then we apply delayed checks using the time to failure statistical property discussed in Section~\ref{sec:spa-for-optimus} and compare against a baseline Optimus implementation.

The relevant Optimus implementations are as follows:

\begin{itemize}
    \item \textbf{Baseline}: Static node timeout intervals. This scenario provides baseline availability and calibration overhead metrics.
    \item \textbf{High-Frequency}: Very short static node intervals. Used to gather fine-grained data for analysis by SPAQ.
    \item \textbf{SPAQ-Informed}: An adaptive version of Optimus with delayed node timeout intervals after calibration, where length of delays is informed by SPAQ analysis of the high-frequency data.
\end{itemize}

\begin{figure}
  \centering
  \includegraphics[width=\linewidth]{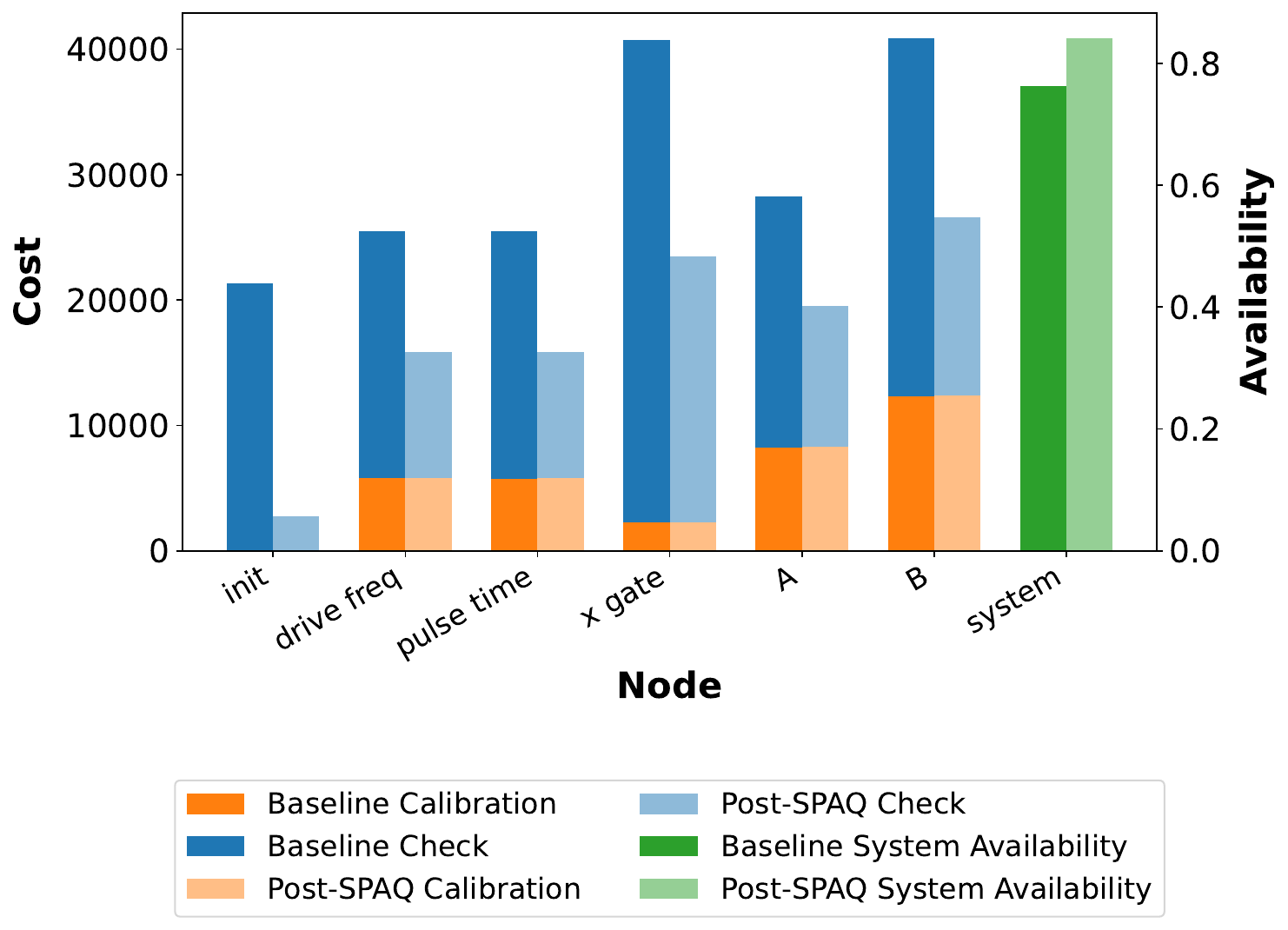}
  \caption{Calibration cost and system availability is analyzed for Experiment 1. Cost is defined as time spent by each node on check and calibration operations. System availability increases from 76.2\% to 84.1\% when moving from Baseline Optimus to our SPAQ-informed adaptive Optimus implementation.} 
  \label{fig:ttf-availability}
\end{figure}

SPAQ analyzes data from the high-frequency implementation and uses the results to inform our Optimus scheduler, increasing the node's timeout value by delaying \texttt{check\_data} operations immediately following calibration (the time when the node is most likely to be in spec).
We compare the total time spent by each node on checks and calibrations, as well as the total system availability between the baseline and SPAQ-informed implementations. Results are found in Fig.~\ref{fig:ttf-availability}.

As expected, calibration overhead for each node remains consistent due to similar failure rates.
However, SPAQ-informed Optimus significantly reduces unnecessary checks by delaying initial post-calibration checks when the node is assumed to be within spec. This results in increased system availability, improving from 76.2\% in the baseline to 84.1\% in the adaptive implementation.

SMC is a tool best used for detecting fine-grained statistically significant differences to the hundredth of a percent, which is not the case when dealing with long-lived calibration nodes. However, this experiment still serves as a clear and practical example of how SPA can be used to extract and apply a meaningful metric.
In the following experiments, we consider more sophisticated hypothesis design.

\subsection{Experiment 2: Inter-Node Parameter Compensation}
Calibration experiments can be tightly interconnected, with some parameters strongly influenced by others.
When this is the case, calibrating one node may shift its parameter and therefore lead to failure of another node.
For example, in an ion trap, voltage compensation may shift ion position relative to a laser beam, affecting optimal gate time.
If situations such as this are not well-managed, failure will go undetected until the dependent node is checked again, leading to unnecessary downtime. 
Property definition using temporal logic is particularly adept at these descriptions, as it is valid and easy to specify events that occur at different points in time and perform statistical tests on them.

\begin{figure}
  \centering
  \includegraphics[width=0.75\linewidth]{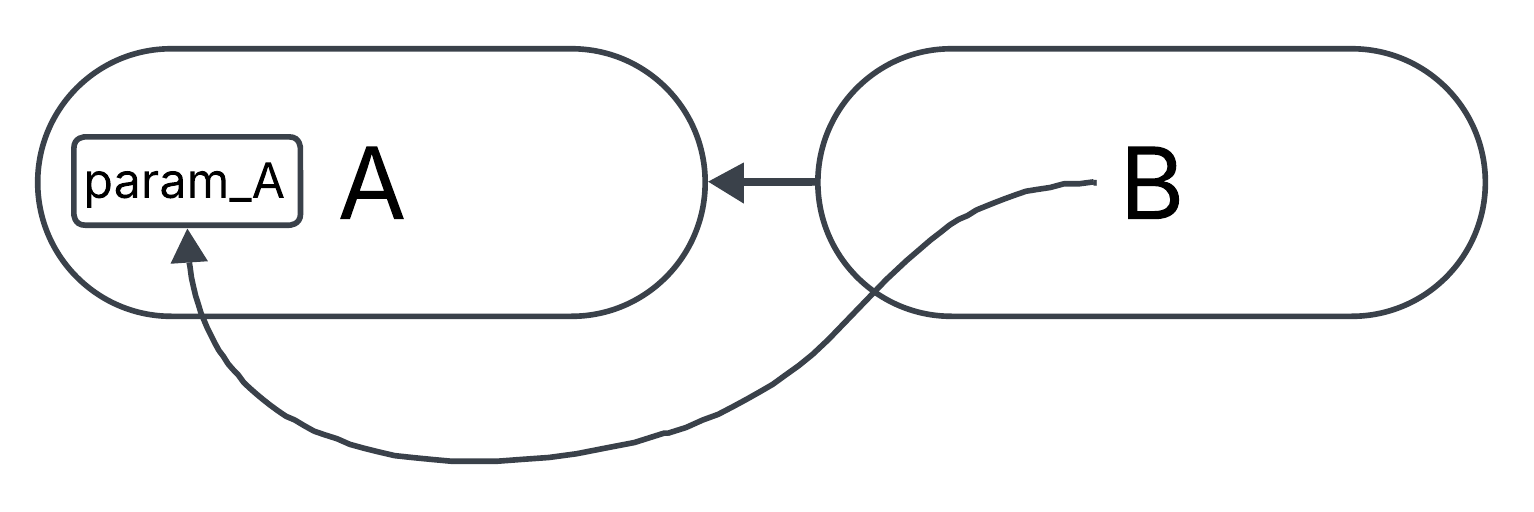}
  \caption{Zoomed-in view of nodes A and B from the calibration graph in Fig.~\ref{fig:optimus-dag-sim} as simulated in Experiment 2. Node B directly depends on a parameter in A, where large changes result in node B failure.} 
  \label{fig:internode-parameter-graph}
\end{figure}
\begin{figure}
  \centering
  \includegraphics[width=0.9\linewidth]{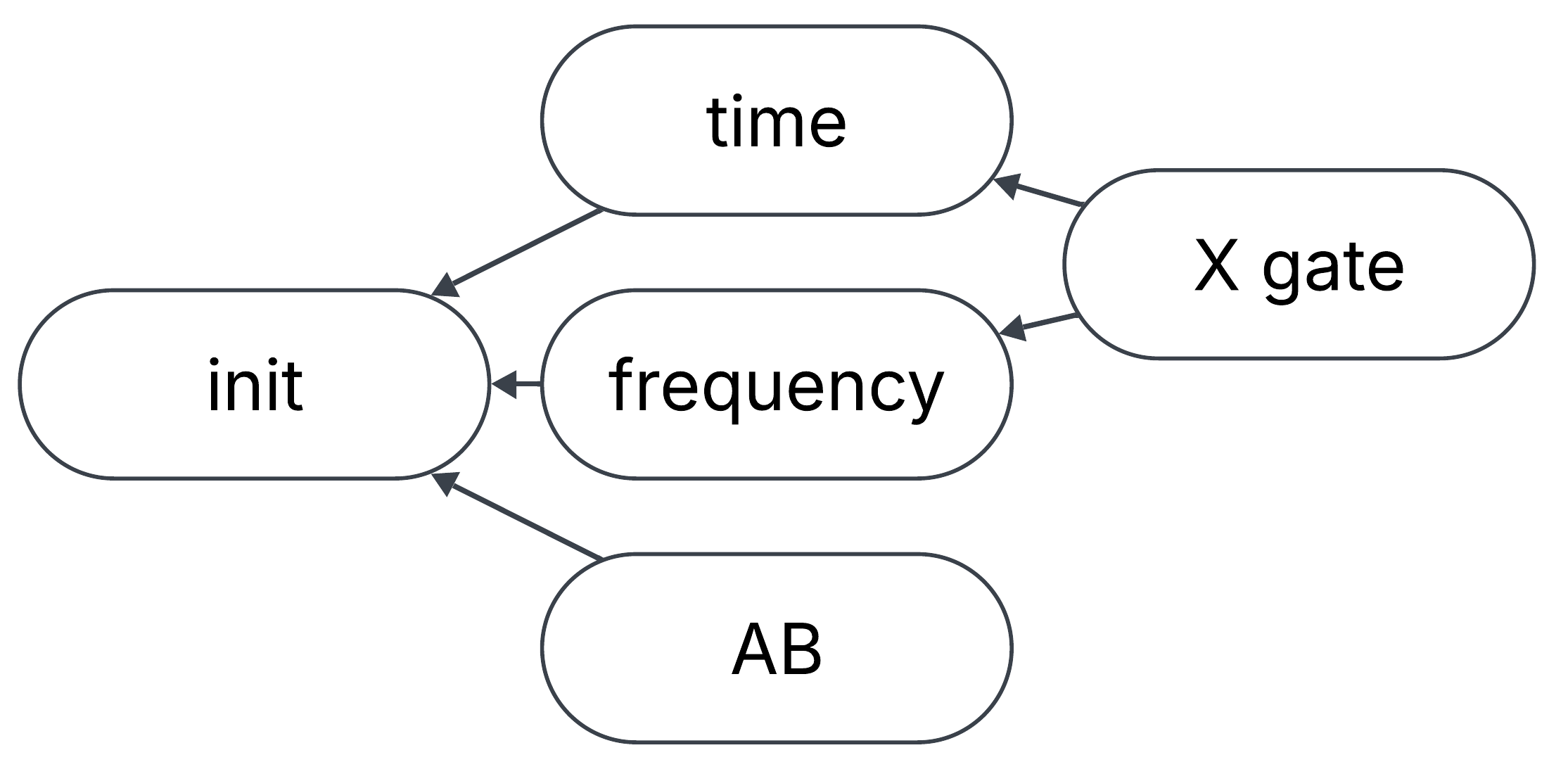}
  \caption{Calibration graph for Experiment 2 after combining nodes A and B into a single node AB.} 
  \label{fig:combined-graph}
\end{figure}
In this experiment, we test the hypothesis, \emph{`A shift of more than 10\% in parameter value A will cause node B to fail next time it is checked with at least a 33\% chance and 95\% confidence.'}
We run Optimus on nodes A and B in Fig.~\ref{fig:optimus-dag-sim}, simulated in isolation from the rest of the graph structure.
We assume these nodes to be complex, potentially with multiple parameters.
We also note that system-wide availability gain is diminished when accounting for downtime from excluded nodes.
This experiment differs from the hidden node scenario of Experiment 3, where we explore whether correlated node failures are caused by a hidden dependency. 
SPAQ finds that node B failures are caused by changes in node A's parameter \emph{param\_A}, and not by generic failure of node A.
This dependency is illustrated in Fig.~\ref{fig:internode-parameter-graph}.


This positive result prompts an informed decision about how to best prevent system downtime, knowing that when parameters in node A drift significantly, node B will likely fail.
We may combine the two nodes into one--this approach is most useful if the two nodes are already connected by an edge, do not disturb the rest of the graph by combining, and their combined \texttt{check\_data} function is easy to perform. 
Alternatively, we could implement custom calibration calls, an approach that works best when two nodes must remain separate: in the case that \emph{param\_A} was found to have shifted by a large amount, an automatic call to calibrate node B could be issued. 
SMC can also help to identify the amount that parameter A must shift to cause failure in node B, further informing actions taken to optimize the calibration graph.

\begin{figure}
  \centering
  \includegraphics[width=\linewidth]{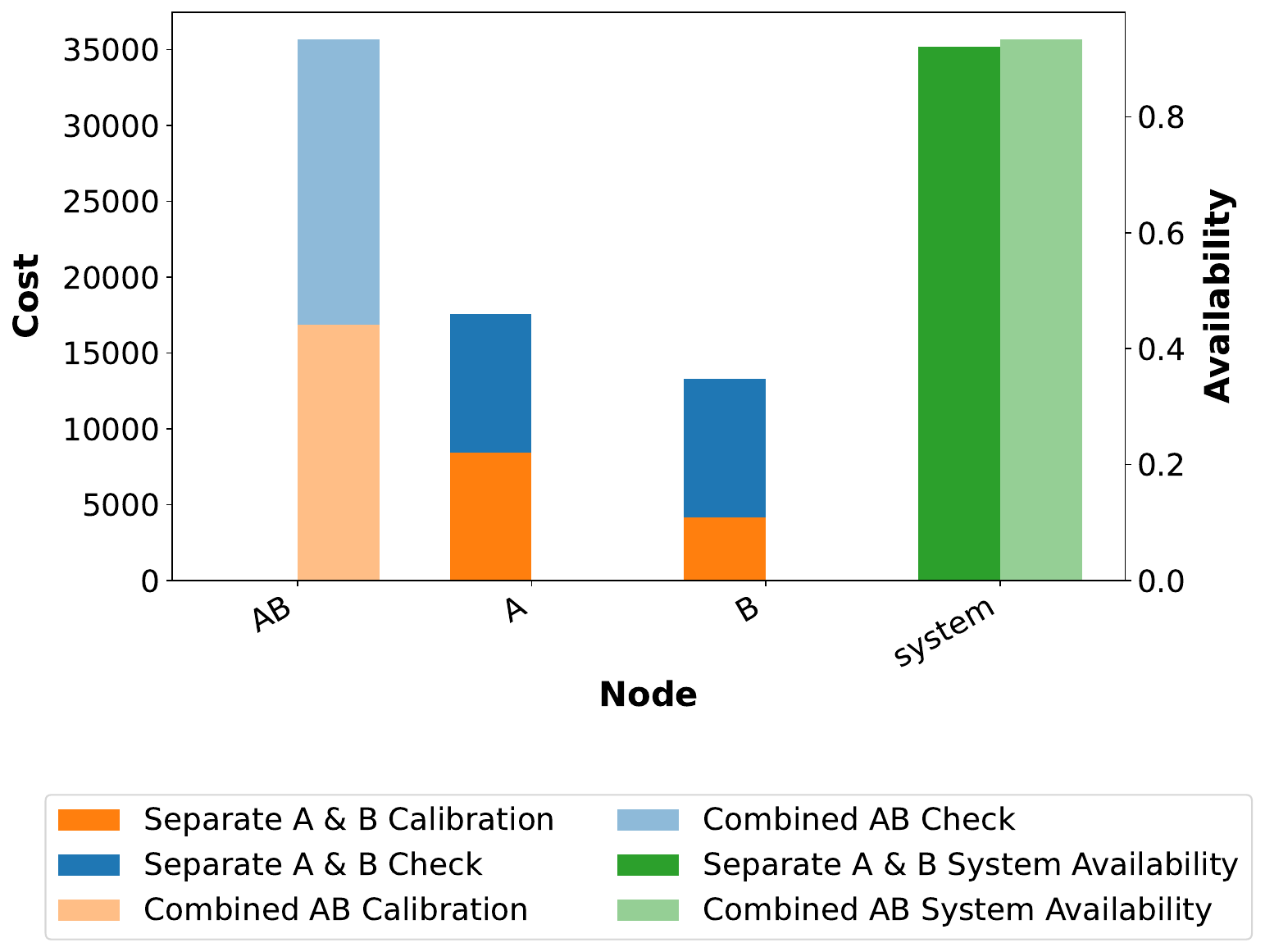}
  \caption{Per-node cost and system availability result for Experiment 2. System availability increases from 92.2\% to 93.4\% after updating the dependency graph to combine nodes A and B.} 
  \label{fig:combined-availability}
\end{figure}

For this experiment, we decide to combine nodes A and B to create the resulting graph seen in Fig.~\ref{fig:combined-graph}.
A and B are both terminal nodes, and so we find it more optimal to combine the two into a single node.
Combining A and B results in an increase in system availability from 92.2\% to 93.4\% as illustrated in Fig.~\ref{fig:combined-availability}.

\subsection{Experiment 3: Hidden Node Connections}
\begin{figure}
   \centering
   \begin{subfigure}{\linewidth}
       \centering
       \includegraphics[width=\linewidth]{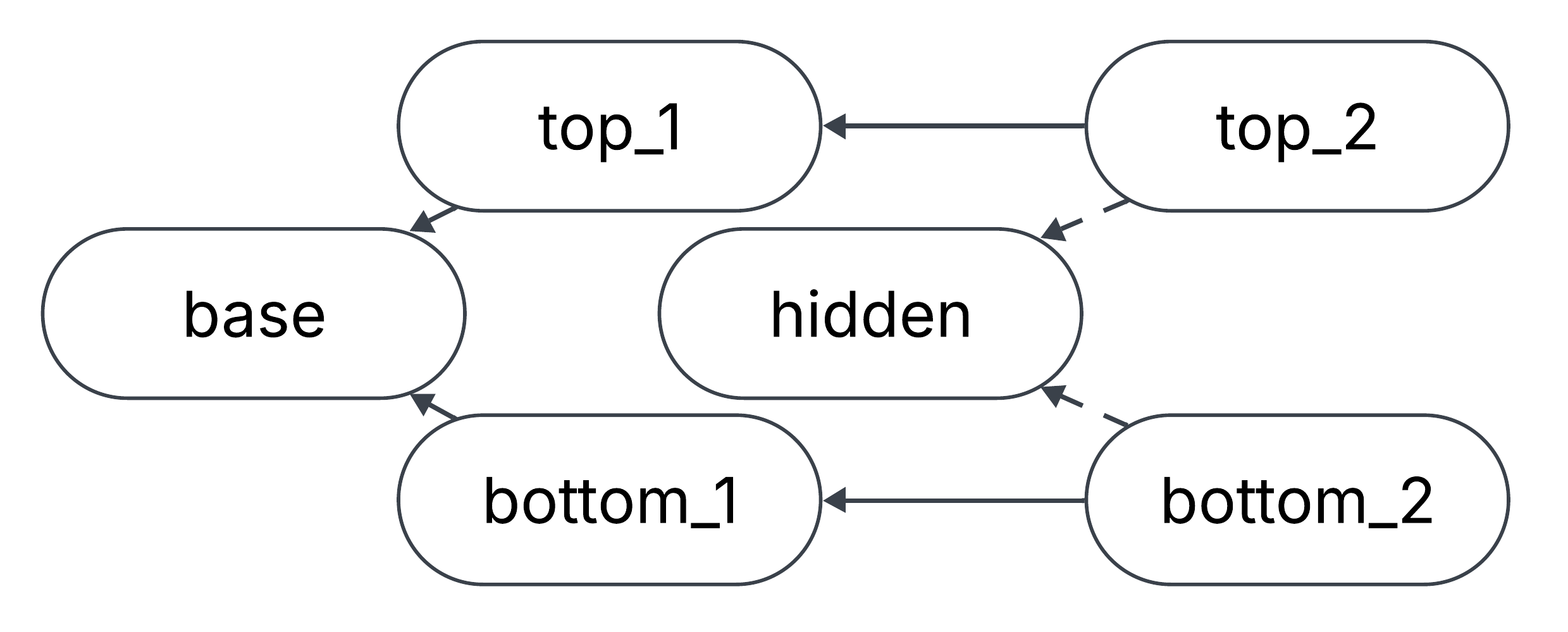}
       \caption{DAG with a hidden dependency shared by two nominally independent nodes, representing indirect or environmental dependencies.} 
       \label{fig:hidden-node-graph}
       \end{subfigure}
   \begin{subfigure}{\linewidth}
       \centering
      \includegraphics[width=\linewidth]{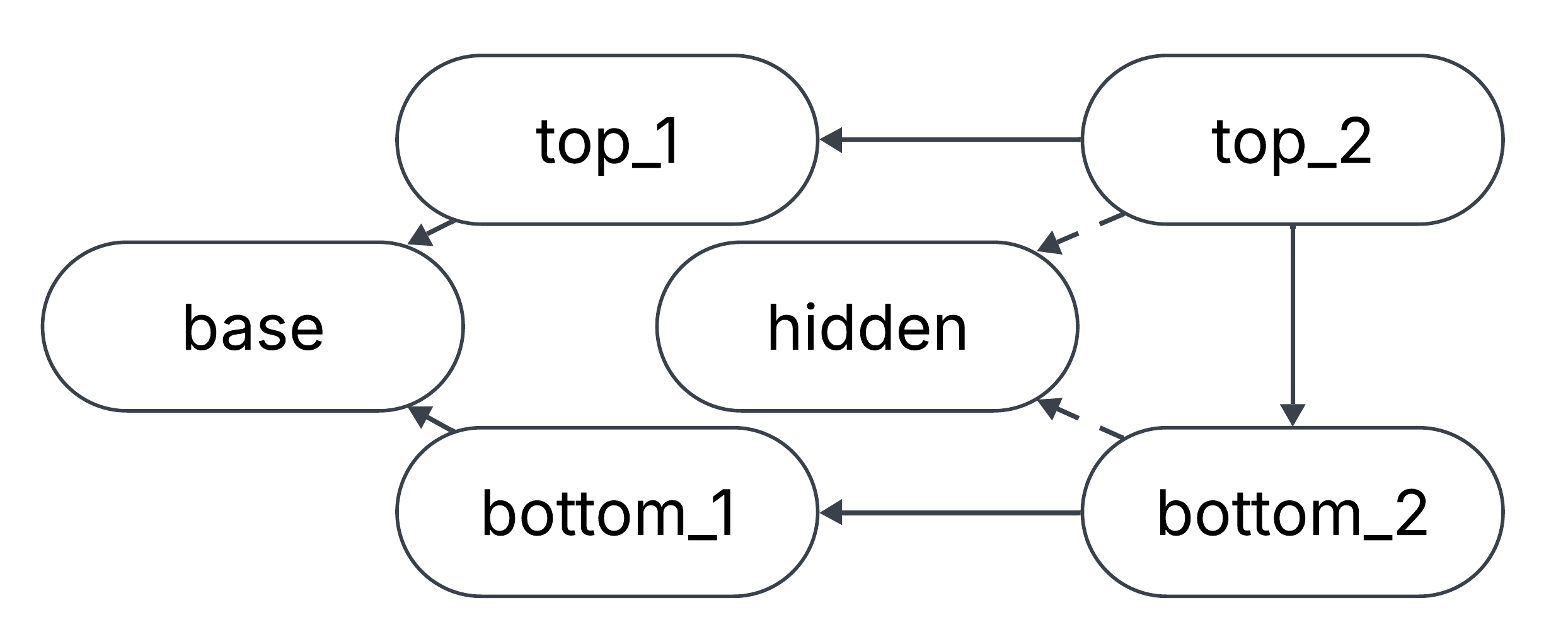}
       \caption{Optimized DAG after the node relationship between \emph{top\_2} and \emph{bottom\_2} is revealed using hypothesis testing.} 
       \label{fig:hidden-node-graph-redraw}
   \end{subfigure}
   \caption{Optimus calibration DAGs for Experiment 3.}
   \label{fig:hidden-node-graphcombined}
\end{figure}

Finally, we perform an experiment to detect hidden dependencies between nodes. We introduce a hidden node between two nominally independent nodes in a dependency graph (nodes \emph{top\_2} and \emph{bottom\_2} in Fig.~\ref{fig:hidden-node-graph}). The hidden node represents a shared dependency such as common control hardware or environmental factors and therefore does not participate in Optimus check and calibration operations.

\begin{figure}
   \centering
   \includegraphics[width=\linewidth]{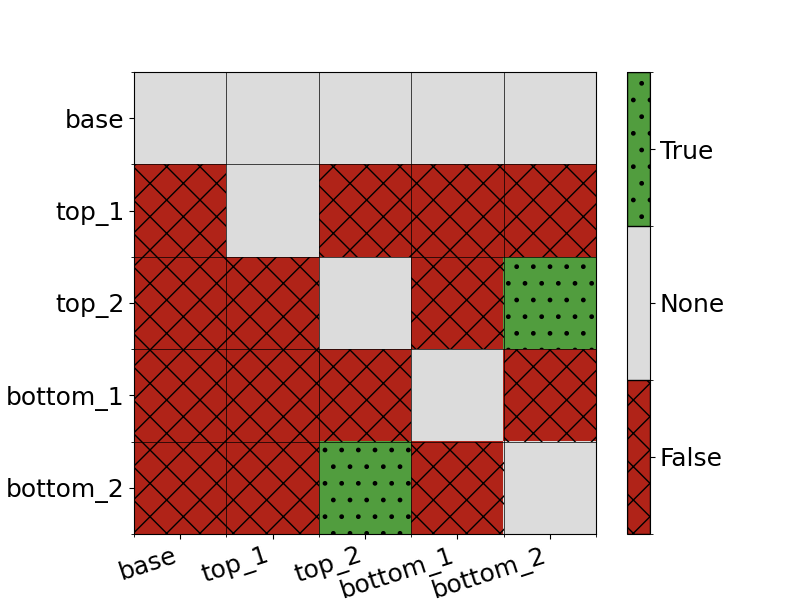}
   \caption{SPAQ significance results of node correlations in Experiment 3.}
   \label{fig:hidden-node-heatmap}
\end{figure}

We perform a hypothesis test using SPAQ: \emph{`Does node Y fail within 25 cycles of node Z with greater than 33\% probability, with 90\% confidence?'}. 
This hypothesis test is applied to every combination of nodes in the graph, where the result for node pair Y and Z is not necessarily the same as for node pair Z and Y. 
Results are shown as a heatmap in Fig.~\ref{fig:hidden-node-heatmap}. 
Most node pairs do not exhibit significant correlation, even if directly connected, which is consistent with independently drifting parameters.
The lack of result between the \emph{base} node and other nodes is attributed to an insufficient number of base node failures to support statistical analysis, as SMC requires a minimum number of samples to reach conclusions.
However, the nominally independent nodes \emph{top\_2} and \emph{bottom\_2} exhibit statistically significant correlation, revealing the hidden dependency.

\begin{figure}
  \centering
  \includegraphics[width=\linewidth]{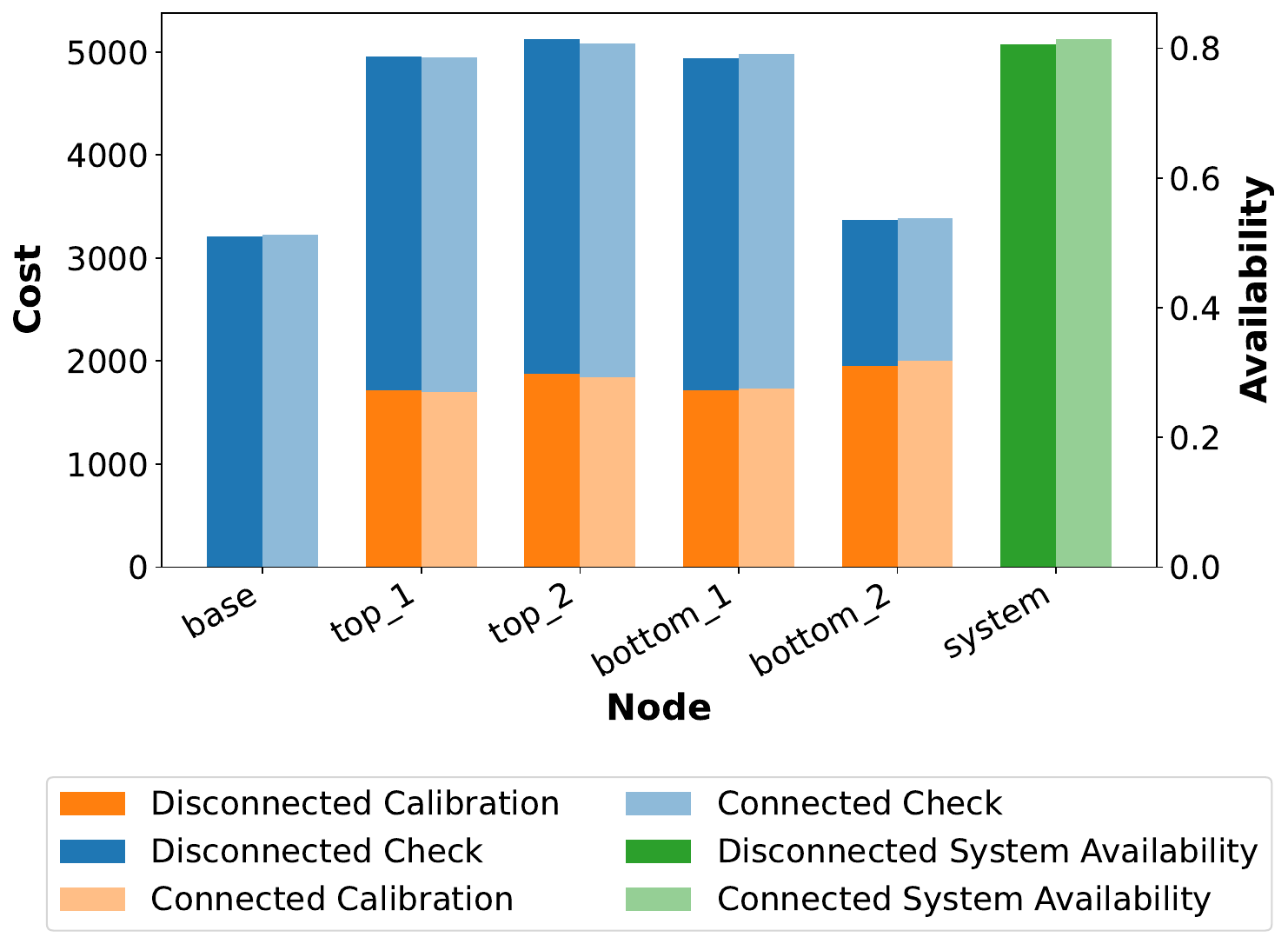}
  \caption{Per node check and calibration cost and overall system availability for Experiment 3, before and after redrawing the DAG to account for the hidden dependency. System availability increases from 80.6\% to 81.4\%.} 
  \label{fig:hidden-availability}
\end{figure}

In order to address this dependency, we add an edge from node \emph{top\_2}, which has a shorter timeout value, to node \emph{bottom\_2}, which has a longer timeout, as shown in Fig.~\ref{fig:hidden-node-graph-redraw}.
We compare system availability before and after modifying the DAG structure, summarizing results in Fig.~\ref{fig:hidden-availability}. Despite an increase in the number of checks, overall system availability increases from 80.6\% to 81.4\%. This demonstrates that identifying and addressing hidden correlations can lead to performance gains.

\section{Future Work}

In this section, we outline several promising avenues for extending the use of SPAQ to enhance quantum systems. We also highlight the compatibility of SPAQ with real quantum devices and encourage its use on operational hardware.

\subsection{Additional Research Directions}
We list several examples of future directions that may further enhance Optimus using the SPAQ framework:

\begin{itemize}
    \item \textbf{Dynamic graph structure}: Optimus calibration graphs are static, but dynamic graph structures could better reflect real-time quantum device usage. For example, if a particular operation appears frequently within queued experiments, its associated node should be kept tightly within spec. Conversely, operations not in use to be temporarily pruned from the calibration graph.
    \item \textbf{Dynamic parameter thresholding}: Thresholds for parameter check data values are static. However, with broader parameter space exploration, it is possible to dynamically adjust thresholds using real-time values in other nodes.
    \item \textbf{SMC property exploration}: 
    This work is limited to metric extraction and pairwise event correlation. Formulating multiple cause-effect relationships between events could potentially determine more subtle properties in quantum system calibration.
    \item \textbf{SMC Hyperproperties}: Allowing expression across multiple executions, as opposed to collections of the same execution, could allow for more robust hypothesis expression. 
    This allows formulation of properties across systems, such as for discovery of hidden nodes that affect X gates on different quantum computers in the same environment~\cite{clarkson:jcs:2010, clarkson:icpst:2014}.
    \item \textbf{SPAQ-Calibration Coupling}: The advanced calibration techniques presented in this work still relay on manual intervention, e.g., to alter thresholds and intervals.
    Tight coupling between SMC and Optimus can allow for immediate, automated informed calibration adjustment and therefore better system availability.
\end{itemize}

Through these expansions, SPAQ can be applied to perform a wider variety of analyses with greater efficiency, improving near-term quantum computing.
However, SPAQ, and SMC in general, is not limited to analysis of quantum system calibration. 
SMC is currently used in classical computing for verification of critical hardware and software, such as those used in robotics and medical devices, where users may want to show that the system will not be in an error state for more than a millisecond~\cite{arney2010patient,roohi:hscc:2017,kwiatkowska2018probabilistic}.
We envision that we can apply the same techniques for the analysis of modular quantum architectures for statistical guarantees.

\subsection{Adaptation for Real Devices}
\label{sec:real-devices}

Our demonstration of SPAQ uses a simulation based on trapped ion quantum computing. However, its core statistical model checking methods are generalizable across quantum computing platforms and calibration techniques.
Minimal changes are needed to run SPAQ-informed Optimus on real quantum devices- users can simply log Optimus data for analysis.
Our open-source GitHub repository\textsuperscript{1} includes an example of how to save all required information to an influx database.

A challenge on real systems is the time needed to collect enough failures for statistically meaningful results.
For example, using SPAQ to achieve 95\% confidence at the 5th and 95th percentiles requires just over 50 samples. 
For ultra-stable nodes that rarely fail and do not generate enough data for analysis, large timeouts can be set or can be monitored indirectly through downstream nodes.

\section{Conclusions}
\label{sec:conclusions}

In this paper, we introduced SPAQ, a framework which leverages SMC for the analysis of node-based quantum calibration procedures. 
We demonstrated SPAQ's application to Optimus, a node-based quantum system calibration method, by using it to analyze calibration runs and formulate hypothesis using temporal logic.
This allowed us to identify key metrics such as time to failure and make informed adjustments to node timeout values.
Our analysis reveals instances where nodes fail simultaneously due to external factors, providing insight for modifying the calibration DAG.
We also identified cases where parameters in one node influence others, allowing us to adjust the calibration graph structure in a way that increases system uptime.

SPAQ can be extended to broader applications than just graph-based calibration, and we anticipate that future work will consider more sophisticated analysis such as hyperproperty exploration and dynamic thresholding.
The core statistical model checking methods introduced by SPAQ are generalizable across various quantum computing platforms and calibration protocols, and may be applied to real devices with minimal effort.
SPAQ is available on our GitHub repository.\textsuperscript{1}

\section{Acknowledgment}
\label{sec:acknowledgment}

This work was supported by a collaboration between the U.S. DOE and other Agencies. This material is based upon work supported by the U.S. Department of Energy, Office of Science, National Quantum Information Science Research Centers, Quantum Systems Accelerator. This work was also supported by the NSF STAQ project (PHY-2325080) and the LPS/ARO QCISS program (W911NF-21-1-0005).

\sloppy
\theendnotes
\fussy

\bibliographystyle{IEEEtran}
\bibliography{refs}

\end{document}